\documentclass{midl} % Include author names
\usepackage{mwe} % to get dummy images
\usepackage{graphicx}
\usepackage{amsmath}
\usepackage{pgfplots}
\usepackage{pgfplotstable}
\usepackage{multicol}
\usepackage{multirow}
\usepackage{graphics}
\usepackage{booktabs}
\usepackage{caption}
\usepackage{wrapfig}
\usepackage{graphicx}
\usepackage{enumitem}

\newcommand\chex{CheXpert}
\newcommand\mimic{MIMIC-CXR}
\newcommand\VinDr{VinDr-CXR}
\newcommand{\best}[0]{$t_\theta$ }\xspace
\newcommand{\edit}[1]{{#1}}

\jmlryear{2023}
\jmlrworkshop{Full Paper -- MIDL 2023}
\editors{Accepted for publication at MIDL 2023}

\title[Exploring Image Augmentations for Siamese Representation Learning with Chest X-Rays]{Exploring Image Augmentations for Siamese Representation Learning with Chest X-Rays
}

\midlauthor{\Name{{Rogier van der} Sluijs\midljointauthortext{Contributed equally}\nametag{$^{1}$}} \Email{sluijs@stanford.edu}\\
\Name{Nandita Bhaskhar\midlotherjointauthor\nametag{$^{2}$}} \Email{nanbhas@stanford.edu}\\
\Name{{Daniel L} Rubin\nametag{$^{1,3,4}$}} \Email{dlrubin@stanford.edu}\\
\Name{{Curtis P} Langlotz\nametag{$^{1,3,5}$}} \Email{langlotz@stanford.edu}\\
\vspace{0.5em}\Name{{Akshay S} Chaudhari\nametag{$^{1,3}$}} \Email{akshaysc@stanford.edu}\\
\addr $^{1}$ Department of Radiology, Stanford University \\
\addr $^{2}$ Department of Electrical Engineering, Stanford University \\
\addr $^{3}$ Department of Biomedical Data Science, Stanford University \\
\addr $^{4}$ School of Medicine, Stanford University \\
\addr $^{5}$ Department of Biomedical Informatics Research, Stanford University \\
\\ * Equal Contributions, Co-first authors
}

\begin{document}

\maketitle

\begin{abstract}
Image augmentations are quintessential for effective visual representation learning across self-supervised learning techniques. While augmentation strategies for natural imaging have been studied extensively, medical images are vastly different from their natural counterparts. Thus, it is unknown whether common augmentation strategies employed in Siamese representation learning generalize to medical images and to what extent. To address this challenge, in this study, we systematically assess the effect of various augmentations on the quality and robustness of the learned representations. We train and evaluate Siamese Networks for abnormality detection on chest X-Rays across three large datasets (\mimic{}, \chex{} and \VinDr{}). We investigate the efficacy of the learned representations through experiments involving linear probing, fine-tuning, zero-shot transfer, and data efficiency. Finally, we identify a set of augmentations that yield robust representations that generalize well to both out-of-distribution data and diseases, while outperforming supervised baselines using just zero-shot transfer and linear probes by up to 20\%. Our code is available at \url{https://github.com/StanfordMIMI/siaug}.

\end{abstract}

\begin{keywords}
Data Augmentations, Self-Supervised Learning, Medical Imaging.
\end{keywords}

\section{Introduction}
\label{sec:intro} 
Deep learning algorithms enable high-accuracy medical image analysis, yet are constrained by limitations of labelled data. Determining ground-truth image labels for diagnostic and prognostic tasks typically involves multiple annotators with clinical expertise and is often costly, time-consuming, and subject to inter-reader variability \cite{kim2022did}. Such a scarcity of annotated datasets has spurred research in data-efficient deep learning techniques, such as transfer learning  and self-supervision \cite{krishnan2022self}. ImageNet pretraining is common, yet transferring representations from natural images is not always successful, possibly due to the shifted distribution and visual features of medical images \cite{raghu2019transfusion}. Self-supervision, on the other hand, exploits the intrinsic structure of unlabelled data to learn effective representations, which can then be used for fine-tuning or zero-shot transfer on downstream tasks. Self-supervision proves to be particularly useful in medicine, given the abundance of unlabelled imaging data. It also provides robustness to out-of-distribution data \cite{hendrycks2019using} and concept drifts. Learning visual features without a strong supervisory signal, however, is challenging.

One particularly powerful technique used in self-supervision is to compare two or more augmented views of the same image using a Siamese network architecture \cite{bromley1993signature}. A common denominator among variants of this technique, such as contrastive learning \cite{chen2020simple, he2020momentum} and feature prediction \cite{grill2020bootstrap, caron2021emerging, chen2021exploring}, is their reliance on an augmentation strategy to generate different views of the input data. The question ``what makes good views'' has been explored in-depth for natural images in the context of contrastive learning \cite{tian20what, chen2020simple}, but has not been answered for medical tasks. Efforts to transfer common augmentation strategies to pretrain representations on medical data have thus far had limited success compared with hand-crafted strategies \cite{azizi2021big, sowrirajan2021moco}.

To address these limitations, we systematically evaluate the effectiveness, robustness, and generalizability of image augmentation strategies for representation learning on three large datasets of chest x-rays \cite{irvin2019chexpert, johnson2019mimic, nguyen2022VinDr}. In this study, we assess an extensive range of augmentations through linear probing, zero-shot transfer, fine-tuning, and data efficiency experiments and show that: 

\begin{itemize}
\item Visual representations extracted with different augmentations results in substantial variations on downstream classification tasks (up to 18\% difference). Random resized cropping largely defines optimal performance of the learned representations on downstream tasks.
\item Representations learned with the optimal set of augmentations outperform supervised baselines on several occasions on both internal (by 13.6-20.0\%) and external \edit{evaluation} (up to 27.0\%) 
 sets. 
 \item Zero-shot transfer, linear probing, and fine-tuning with limited data using pretrained representations surpass classification accuracy of their supervised counterparts on several occasions.
\item The learned features are robust to forms of label drift and catastrophic forgetting, and show success in classification of diseases that are rare \edit{(e.g. Rib Fractures [RF])} and unseen across datasets (e.g. \edit{Tuberculosis [TB]}). 
\end{itemize}

\section{Related Work}
\label{sec:related_work}
\textbf{Self-supervised learning}.
Self-supervision typically involves formulating a pretext task solely to learn a good representation of the data. This representation can subsequently be fine-tuned on a downstream task in a data-efficient manner. A broad range of such pretext tasks exist, such as solving jigsaw puzzles \cite{noroozi2016unsupervised, taleb2021multimodal}, image rotation prediction \edit{\cite{gidaris2018unsupervised}}, and context restoration \cite{pathak2016context}. 

\vspace{2pt}
% \noindent 
\textbf{Contrastive learning}. 
Contrastive visual representation learning seeks to contrast positive pairs of image views with negative pairs \cite{hadsell2006dimensionality}. Positive pairs are created from the input data, whereas negative pairs are sampled from a mini-batch \cite{chen2020simple} or queue \cite{chen2021exploring}. Traditional contrastive learning requires positive pairs and a large sample of negative pairs for effective training. Variations of contrastive methods use approaches that do not rely on negative pairs. BYOL \cite{grill2020bootstrap} introduced a Siamese network trained to predict views of opposing branches. Extensions of this framework explore different architectural components, such as the loss function, projection heads, and the teacher-student architecture \cite{caron2021emerging, chen2021exploring}.
\vspace{2pt}

% \noindent 
\textbf{Image augmentations for self-supervision}.
% How natural image augmentations are used
Data augmentations are widely used in supervised learning to increase the diversity of the training data and to improve generalizability \cite{krizhevsky2017imagenet, cubuk2018autoaugment}. RandAugment \cite{cubuk2020randaugment} is a powerful method that applies a randomly selected subset of predefined augmentations to the input data. In contrast, in self-supervised learning, augmentations are often applied to construct a pretext task \cite{tian20what}. Common augmentations for contrastive learning were explored in SimCLR \cite{chen2020simple}. In the medical domain, amongst others, affine transformations, elastic deformations \cite{chaitanya2020contrastive}, and physics-driven augmentations \cite{desai2021vortex} have been considered for self-supervised learning.

\vspace{0.2em}
% \noindent 
\textbf{Self-supervised learning for Chest X-Rays}.
Chest X-Ray classification is a well-studied subject, and its recent role has been amplified in light of the COVID-19 pandemic \cite{wynants2020prediction}. Self-supervision has emerged as viable strategy to aid the detection of pathologies on chest x-rays \cite{gazda2021self, azizi2022robust}. Multi-modal vision-language learning has shown to be effective \cite{zhang2020contrastive,huang2021gloria,tiu2022expert,delbrouck-etal-2022-vilmedic}, but necessitates the availability of radiology reports. The current study is most closely aligned with the image-only augmentation strategies examined in MoCo-CXR \cite{sowrirajan2021moco}, MICLe \cite{azizi2021big}, \edit{and REMEDIS \cite{azizi2022robust}}. These studies, however, use contrastive methods that rely on negative sampling and were not designed to systematically explore augmentation strategies. 

\section{Methods}
\label{sec:methods} 
To evaluate the impact of data augmentations on the quality of the learned representations, we used SimSiam \edit{\cite{chen2021exploring}} - a minimal Siamese network architecture. SimSiam does not rely on negative sampling, knowledge distillation, or prototype clustering, which allows us to most directly study the role of augmentations in Siamese representation learning. 

\subsection{Architecture and Pretraining Objective}
\label{sec:architecture}
\begin{wrapfigure}{l}{0.35\textwidth}
\includegraphics[width=0.35\textwidth]{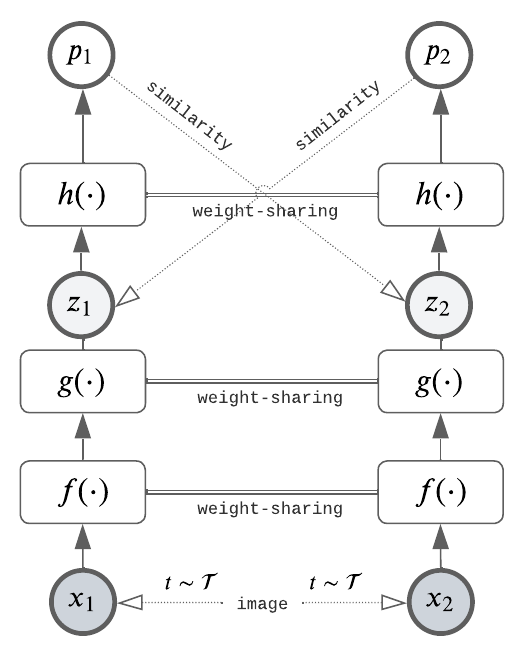} 
\caption{\small SimSiam architecture.}
\label{fig:architecture}
\end{wrapfigure}

The architecture of SimSiam consists of two identical and weight-sharing branches that each take an augmented view (i.e. $x_1$ and $x_2$) of the same image $x$ as an input (\figureref{fig:architecture}). Both views ($x_1$ and $x_2$) are processed by an identical encoder network, $f(\cdot)$, that outputs feature vectors $f(x_i)$. These feature vectors are passed on to a two-layered MLP projector network $g(\cdot)$ that produces a low-dimensional latent representation $z_i = g(f(x_i))$ of the data. 
As a final step, the latent representations produced by each branch ($z_i$) are input to a predictor network $h(\cdot)$. The predictor network is an MLP that aims to predict the projection $z$ of the opposing branch (i.e. $h_1(z_1) = p_1$ tries to predict $z_2$, while $h_2(z_2) = p_2$ tries to predict $z_1$). The loss function $\mathcal{L}$ is defined as the negative cosine similarity between the predictions of the predictor networks $p_1$ and $p_2$ and the actual projected feature vectors $z_1$ and $z_2$:

\begin{equation}
\mathcal{L} = 
-\frac{1}{2} \left(\frac{p_1}{\left\| p_1 \right\|_{2}} \cdot \frac{z_2}{\left\| z_2 \right\|_{2}}\right) 
-\frac{1}{2} \left(\frac{p_2}{\left\| p_2 \right\|_{2}} \cdot \frac{z_1}{\left\| z_1 \right\|_{2}}\right)
\end{equation}

\vspace{8pt}
\noindent 
where $\left\| \cdot \right\|_{2}$ is the $l_2$ norm. Note that, unlike typical contrastive self-supervised learning, calculation of the loss does not involve negative samples.\\
\\
%\vspace{3em}
\vspace{-3mm}
\subsection{Data Collection}
\label{sec:data_collection}
Frontal chest x-rays from three publicly available datasets were used to train and evaluate our models. First, the \mimic{} \cite{johnson2019mimic} dataset (from Boston, USA) includes images acquired from 277,835 imaging studies of patients, of which \edit{$n=200,000$ images were used for training and validation, and $n=37,962$ were used for evaluation}. 
Second, the \chex{} \cite{irvin2019chexpert} dataset (from Stanford, USA) contains chest x-rays from 65,240 patients, of which \edit{$n=168,660$ images were used for training and validation, and $n=22,367$ were used for evaluation.} 
In both \chex{} and \mimic{}, an automatic radiology report labeller \cite{irvin2019chexpert}  was used to annotate each report/image pair for the presence of 14 different conditions of which a diverse subset was included (Appendix \ref{app:data_dist}, Table \ref{tab:chex_mimic_distribution}). Third, the \VinDr{} \cite{nguyen2022VinDr} dataset (from Vietnam) contains 18,000 chest x-rays of which 15,000 were each manually labelled by three radiologists for 22 critical findings and 6 diagnoses in the training set. Every \edit{evaluation} set image ($n=3,000$) was annotated by five radiologists. The sophisticated labelling makes  \VinDr{} an optimal dataset for evaluation purposes (Appendix \ref{app:data_dist}, Table \ref{tab:vindr_data_distribution}).

% \vspace{-3mm}
\subsection{Experimental Setup and Study Design}
\label{sec:experimental_setup}

\textbf{Training Pipeline}. Our training pipeline consists of (i) self-supervised pretraining of an encoder, $ResNet(\cdot)$, using unlabelled images via SimSiam (Section \ref{sec:architecture}), (ii) supervised linear probing (i.e. training a single-layer classifier on top of a frozen encoder), and (iii) supervised fine-tuning of the entire encoder initialized with the weights of a pretrained encoder and a pretrained classification head. 

\vspace{0.5em}
% \noindent 
\textbf{Datasets}. We use the unlabelled \mimic{} training split for all self-supervised pretraining experiments. We provide pretraining results with \chex{} in Appendix \ref{app:add_results_pairwise_aug} (Tables \ref{tab:chex_pairwise} and \ref{tab:class_wise_chex}). We perform supervised linear probing and fine-tuning on labelled train splits of \mimic{}, \chex{}, and \VinDr{}. For evaluation, we use held out data from the internal \edit{evaluation} split (\mimic{}) as well as external \edit{evaluation} splits (\chex{} and \VinDr{}). All dataset splits and the included labels are detailed in Appendix \ref{app:data_dist} (Tables \ref{tab:chex_mimic_distribution} and \ref{tab:vindr_data_distribution}). 
The three datasets encompass multi-site data and include different diseases and occurrence rates. Some labels are overlapping, while others are unseen in the pretraining data (e.g. \edit{tuberculosis [TB]} in \VinDr{}), forming a test-bed for comprehensive evaluations.

% \vspace{0.2em}
%\noindent 
\textbf{Training Details}. We use the SimSiam architecture \cite{chen2021exploring} with a ResNet-50 encoder \cite{He2016} for all experiments involving representation learning. The SGD optimizer (with $weight\:decay=0.0001$ and $momentum=0.9$) is used for pretraining ($lr=0.05$), linear probing ($lr=30$, $weight\:decay=0$), and fine-tuning ($lr=0.00001$); using a cosine decay learning rate scheduler \cite{chen2021exploring}. Batch sizes were fixed to 256.  
Experiments were trained with PyTorch on 8 NVIDIA A100 GPUs on a single node with 32-bit floating point precision. 

% \vspace{0.2em}
% \noindent 
\textbf{Evaluation Metrics}. We evaluate the quality of our pretrained representations by measuring their downstream discriminative performance (averaged and per label), generalization capability, and data efficiency using the following multi-label metrics: (i) \edit{M}acro AUROC (area under receiver operating curve), (ii) label-wise AUROC, (iii) Hamming Loss, and (iv) Ranking Error \cite{Tsoumakas2010_multilabel} \edit{for holistic evaluation}. We report Macro AUROC in the main manuscript and the rest, \edit{along with their descriptions and motivation} in Appendix \ref{app:metrics}, \ref{app:add_results}.

\begin{figure}
\centering
{\includegraphics[width=0.97\textwidth]{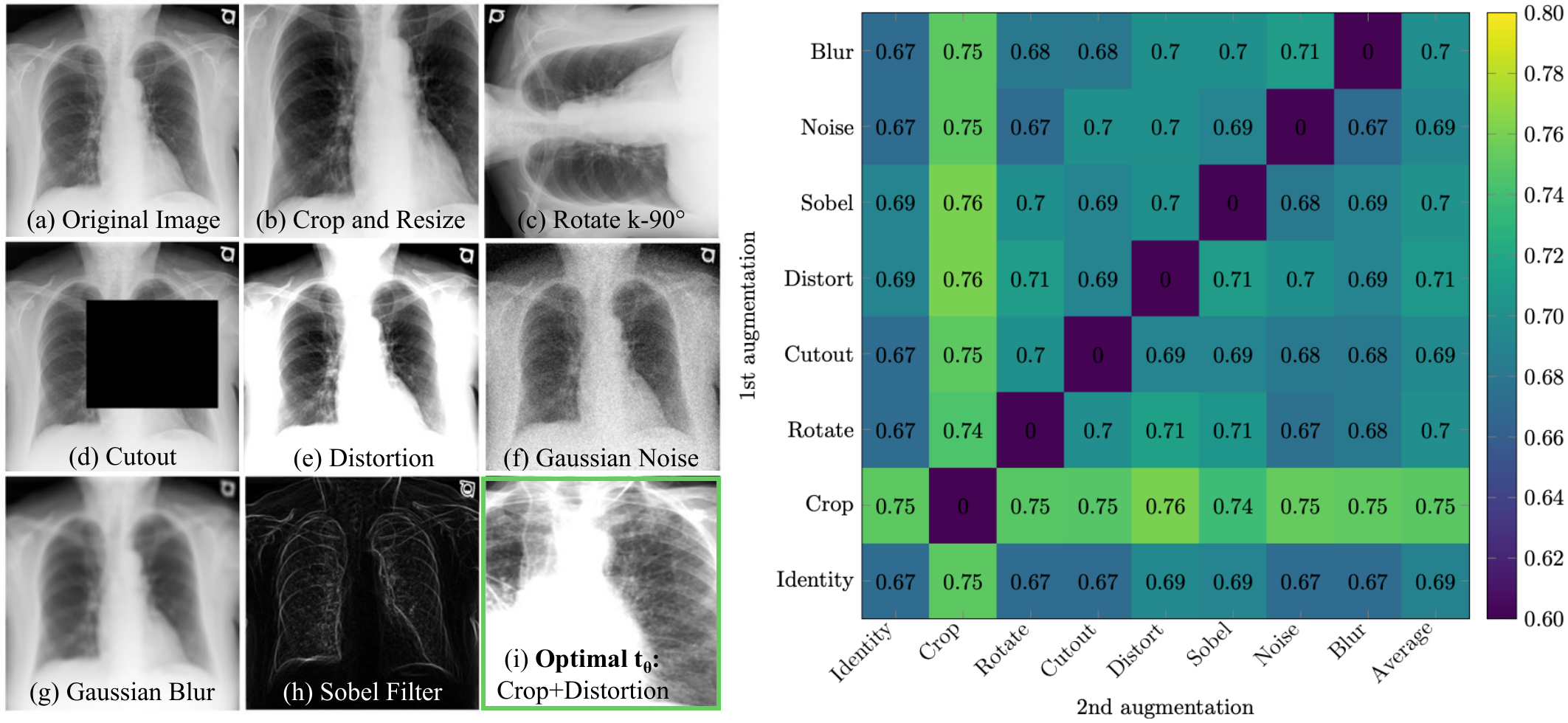}}
\caption{\small A selection of image augmentations (left: a-h) and their performance on \mimic{} \edit{evaluation} data (right), following pretraining and \edit{linear probing} on \mimic{} training data. Combining random resized cropping with distortion resulted in the best augmentation pair, \best.}
\label{fig:augmentations}
\end{figure}

\section{Experiments and Findings}
\label{sec:results}

\subsection{Optimal Augmentation Strategy} 
\label{sec:optimal_aug_strategy}
We seek to learn invariant features from augmented image views during pretraining. Inspired by the systematic study of augmentations for SimCLR by \citet{chen2020simple}, we first explore the efficacy of augmentations in isolation. We evaluate three common geometric/spatial transformations, namely resized cropping, rotation \cite{gidaris2018unsupervised}, and cutout \cite{devries2017improved}, along with pixel-wise transformations of \edit{distortion (i.e. brightness/contrast)}, Gaussian noise, Gaussian blur, and Sobel filtering (\figureref{fig:augmentations}).

First, we apply the identity transformation to one branch of the Siamese network, and apply a single augmentation $t_i \in \mathcal{T}$ to the other branch (i.e. $t_1(x_i)$). We repeat this procedure with pairs of augmentations (i.e. $t_2(t_1(x_i))$) as shown in \figureref{fig:augmentations}. We pretrain our models on \mimic{} and evaluate their performance based on supervised linear probing \cite{zhang2016colorful} on the \mimic{} validation set. We refer to the pair of augmentations with the highest Macro AUROC on the \mimic{} validation set as $t_\theta$. 

We find a combination of random resized cropping and brightness/contrast adjustments (i.e. pixel distortion) to be the optimal pair of augmentations $t_\theta$ with an AUROC of 0.760 (\figureref{fig:augmentations}). Pairs of augmentations that include random resized cropping consistently outperform other compositions (AUROC improvements ranging from \edit{$0.023-0.092$}). This in contrast to natural images, in which cropping performs well, but mostly in conjunction with either color jittering or Sobel filtering \cite{chen2020simple}. We further optimize the hyperparameters of $t_\theta$ and find that strong cropping ($scale = 0.2-0.5$) and large brightness/contrast distortions ($\lambda = 0.7$) are favored for single-branch augmentations, while weaker cropping ($scale = 0.3-0.9$) is favored for symmetrical dual-branch augmentations. A strategy without any augmentations yields surprisingly good results \edit{($0.668$ AUROC) and serves as a baseline to compare other augmentations with}. We report the Hamming Loss and Ranking Error for each of the augmentation pairs for \mimic{} and \chex{} in the Appendix (Tables \ref{tab:mimic_pairwise} and \ref{tab:chex_pairwise}) and observe consistent trends.

Finally, we compare \best head-to-head with RandAugment and observed that RandAugment linear probing on \mimic{} was effective (AUROC of 0.76) but not superior to the simpler \best strategy. We examined \edit{adding a third augmentation as well as several less common augmentations, but do not consider them for further experiments (Appendix \ref{app:appendix_aug})}.

% \vspace{-3mm}
\subsection{Comparison to Fully Supervised Networks}
\label{sec:compare_fs}
We compare \edit{linear probing} performance  \edit{on} $t_\theta$ with fully supervised models trained from scratch [FS (S)] and with ImageNet pretrained weights [FS (IN)] \edit{on \mimic{}} (Table \ref{tab:class_wise_supervised}) \edit{and \VinDr{} (Table \ref{tab:lcls_VinDr})}. We observe that the linear probe on $t_\theta$ (or any pair of augmentations with resized cropping, see Appendix \ref{app:add_results}) surpasses both fully supervised networks for \mimic{} \edit{(by 0.050 and 0.018 AUROC}) and \VinDr{} \edit{(by 0.099 and 0.057 AUROC)}.
Further stratifying the results, we show that the \edit{linear probes} outperform the fully supervised approaches for almost all conditions, including the challenging minority class of \edit{rib fractures [RF]} that has $<5\%$ prevalence, by 0.089 and 0.055 AUROC (Table \ref{tab:class_wise_supervised}).

% \usepackage{booktabs}
\begin{table}[ht]
\captionsetup{skip=-8pt}
\scriptsize
\caption{ \small Comparison of \best to fully supervised models on \mimic{}. Numbers are AUROC. Abbreviations: FS: Fully supervised, S: trained from scratch, IN: ImageNet, AT: Atelectasis, CM: Cardiomegaly, RF: Rib fracture, PE: Pleural Effusion, PNA: Pneumonia, PTX: Pneumothorax.}
\label{tab:class_wise_supervised}
\begin{center}
\begin{tabular}{@{}llllllllll@{}}
\toprule
Strategy & AT & CM & Edema & RF & PE & PNA & PTX & \begin{tabular}[c]{@{}l@{}}No \\ Finding\end{tabular} & \begin{tabular}[c]{@{}l@{}}\edit{M}acro \\ AUROC\end{tabular} \\ \midrule
\edit{Linear probe} & \textbf{0.750} & \textbf{0.769} &\textbf{0.848} &\textbf{0.619} & 0.845 & 0.649 & \textbf{0.779}& \textbf{0.819 }& \textbf{0.760} \\
\edit{Fine tune} & \edit{\textbf{0.799}} & \edit{\textbf{0.806}} & \edit{\textbf{0.880}} & \edit{\textbf{0.678}} & \edit{\textbf{0.892}} & \edit{\textbf{0.725}} & \edit{\textbf{0.841}} & \edit{\textbf{0.856}} & \edit{\textbf{0.810}} \\
FS (S) & 0.705 & 0.720 & 0.797 & 0.529 & 0.837 & 0.617 & 0.689 & 0.786 & 0.710 \\
FS (IN) & 0.744 & 0.753 & 0.832 & 0.563 & 0.856 & 0.667 & 0.748 & 0.770 & 0.742 \\ \bottomrule
\end{tabular}
\end{center}
\end{table}

% \vspace{-5mm}
\subsection{Zero-shot Generalization of Pretrained Representations}
\label{sec:zero_shot}
We evaluate zero-shot transfer of our \best representations to \VinDr{} and \chex{}, which have differing disease distributions and dataset statistics. In zero-shot transfer to \VinDr{} pathologies  available in \mimic{}, the \best  representations achieve \edit{0.805} AUROC, outperforming fully supervised \VinDr{} networks by \edit{0.094} and \edit{0.002} AUROC when trained from scratch and ImageNet, respectively (Table \ref{tab:lcls_VinDr}). This was striking as the pretrained \best representations did not have any access to \VinDr{} data or label distributions. However, such an effective zero-shot transfer was not the case with \chex{}. Here, the fully supervised \chex{} performance was higher than that of the \best representations \edit{(\tableref{tab:generalization_to_chexpert}, \textit{Eval on \chex{}})}. We attribute this zero-shot discrepancy due to the substantially higher amount of labelled images in \chex{} than in \VinDr{} (168,660 vs 18,000). 

\noindent \edit{Furthermore, the linear probes achieve better zero-shot generalization compared to fully supervised models, trained on \mimic{} from scratch 
%or from ImageNet pretrained weights 
(Appendix \ref{app:add_results_zero_shot_comp}, Table \ref{tab:zero_shot_compare}).}

% \vspace{-2mm}
\subsection{\edit{Transferring Pretrained Representations} to \VinDr{}}
\label{sec:transfer_to_vindr}
Here, we linearly probe the \best representations \edit{pretrained on \mimic{}} using the \VinDr{} dataset, which consists of seen and unseen pathologies.  
When evaluating the \mimic{} \best pretrained classifiers on held out \VinDr{} data, we observe 0.099 AUROC and 0.067 AUROC improvements over fully supervised models (same trends for \chex{} representations) in \tableref{tab:lcls_VinDr}. This trend is consistent across \edit{almost} all pathologies \edit{(we report additional results on different data splits in Appendix \ref{app:data_dist}, Table \ref{tab:vindr_balanced_imbalanced})}. Remarkably, the performance on \edit{Tuberculosis [TB]}, an unencountered disease with very low prevalence in the US, has 0.127 and 0.101 AUROC better performance than the supervised baselines (\tableref{tab:lcls_VinDr}). This shows that linear probing of strong pretrained representations can generalize to out-of-distribution, unseen data and pathologies. 

\begin{table}[ht]
\captionsetup{skip=4pt}
\scriptsize
\caption{\small Transferring \best to \VinDr{} for seen (in-distribution) and unseen (out-of-distribution, OOD) conditions in \edit{\mimic{}}. Numbers are AUROC. Abbreviations: FS (S), FS (IN): Fully supervised from scratch and from ImageNet, respectively, CM: Cardiomegaly, PE: Pleural Effusion, PNA: Pneumonia, PF: Pulmonay Fibrosis, PT: Pleural Thicknening, LO: Lung Opacity, TB: Tuberculosis.
}
\label{tab:lcls_VinDr}
\begin{tabular}{ccccc|ccccc|cc}
\toprule
 & \multicolumn{4}{c|}{In-Distribution Pathologies} & \multicolumn{5}{c|}{Out-of-distribution Pathologies} &  &  \\ \midrule
Strategy & \begin{tabular}[c]{@{}c@{}}CM\end{tabular} & \begin{tabular}[c]{@{}c@{}}PE\end{tabular} & \begin{tabular}[c]{@{}c@{}}PNA\end{tabular} & \begin{tabular}[c]{@{}c@{}}No \\ finding\end{tabular} & \begin{tabular}[c]{@{}c@{}}PF\end{tabular} & \begin{tabular}[c]{@{}c@{}}PT\end{tabular} & \begin{tabular}[c]{@{}c@{}}LO\end{tabular} & \begin{tabular}[c]{@{}c@{}}Mass\end{tabular} & \begin{tabular}[c]{@{}c@{}}TB\end{tabular} & \begin{tabular}[c]{@{}c@{}}Macro \\ AUROC\end{tabular} & \begin{tabular}[c]{@{}c@{}}OOD\\ AUROC\end{tabular} \\ \midrule
Zero-shot  & 0.840  & 0.810   & 0.774  & 0.795  & NA  & NA  & NA   & NA  & NA  & NA  & NA    \\
Linear probe  & 0.909   & 0.822  & 0.785  & \textbf{0.880}  & \textbf{0.720}   & \textbf{0.712}   & 0.651    & 0.648   & 0.776  & 0.767   & 0.701   \\
Fine tune  & \textbf{0.937}   & 0.824  & \textbf{0.790}  & 0.869  & 0.719    & 0.707    & \textbf{0.660}   & \textbf{0.651}  & \textbf{0.802}    & \textbf{0.773}  & \textbf{0.708}  \\
FS (S) & 0.796   & 0.643  & 0.591  & 0.813  & 0.631    & 0.665    & 0.622    & 0.598   & 0.649  & 0.668   & 0.633   \\
FS (IN) & 0.888   & \textbf{0.872}  & 0.672  & 0.778  & 0.631    & 0.694    & 0.613    & 0.571   & 0.675  & 0.710    & 0.637   \\
\bottomrule
\end{tabular}
\end{table}

% \vspace{-2mm}
\subsection{Generalization Capability by Fine-Tuning \mimic{} \best on \chex{}}
\label{sec:transfer_to_chex}
We fine-tune the pretrained \mimic{} \best representations and \mimic{} trained linear classifier on labelled \chex{} training data. We see that the classification AUROC increases on fine-tuning from 0.649 to 0.768 AUROC, even outperforming the fully supervised network trained from scratch (0.757 AUROC) (Table \ref{tab:generalization_to_chexpert}, \textit{Eval on \chex{}}). We then evaluate this \mimic{} \best pretrained and \chex{} fine-tuned model for its zero-shot transfer capabilities to \edit{evaluation} data from \mimic{} and \VinDr{} data (Table \ref{tab:generalization_to_chexpert}). Upon fine-tuning these representations on \chex{}, we wish to assess the presence of catastrophic forgetting or poorer generalization to \mimic{} through our zero-shot evaluations.  However, we see that zero-shot evaluation on \mimic{} continues showing high performance (0.763 AUROC), which still outperforms fully supervised models by 0.053 and 0.021 AUROC, indicating no evidence of catastrophic forgetting. In fact, the performance on \mimic{} \edit{evaluation} data is nearly identical (AUROC difference of 0.003) when fine-tuned on \chex{} or not. Similarly, this fine-tuned model generalizes well to \VinDr{} with 0.810 AUROC, which is 0.142 and 0.091 AUROC higher than fully supervised baselines trained on \VinDr{} (Table \ref{tab:generalization_to_chexpert}, \textit{Eval on \VinDr{}}). \edit{We believe that VinDR, being a relatively small dataset, benefits greatly from pretraining and fine-tuning as compared to fully supervised learning}. 

\begin{table}[ht]
\scriptsize
\caption{\small \mimic{} \best to \chex{} transfer on fine-tuning. Macro AUROC. Abbreviations: ZS: Zero-shot transfer, FT: Fine-tuning, FS: Fully Supervised, S: trained from scratch, IN: ImageNet,  Chex: \chex{}, Mimic: \mimic{}, VinDr: \VinDr{}, Eval: Evaluation.}
\label{tab:generalization_to_chexpert}
\begin{tabular}{cccc|cccc|cccc}
\toprule
\multicolumn{4}{c|}{Eval on \chex{}} & \multicolumn{4}{c|}{Eval on \mimic{}} & \multicolumn{4}{c}{Eval on \VinDr{}} \\ \midrule
ZS & \begin{tabular}[c]{@{}c@{}}FT\\ (Chex)\end{tabular} & \begin{tabular}[c]{@{}c@{}}FS (S)\\ (Chex)\end{tabular} & \begin{tabular}[c]{@{}c@{}}FS (IN)\\ (Chex)\end{tabular} & ZS & \begin{tabular}[c]{@{}c@{}}FT\\ (Chex)\end{tabular} & \begin{tabular}[c]{@{}c@{}}FS (S)\\ (Mimic)\end{tabular} & \begin{tabular}[c]{@{}c@{}}FS (IN)\\ (Mimic)\end{tabular} & ZS & \begin{tabular}[c]{@{}c@{}}FT\\ (Chex)\end{tabular} & \begin{tabular}[c]{@{}c@{}}FS (S)\\ (VinDr)\end{tabular} & \begin{tabular}[c]{@{}c@{}}FS (IN)\\ (VinDr)\end{tabular} \\ \midrule
0.649 & 0.768 & 0.757 & \textbf{0.789} & 0.760 & \textbf{0.763} & 0.710 & 0.742 & 0.765 & \textbf{0.810} & 0.668 & 0.719 \\ \bottomrule
\end{tabular}
\end{table}

% \vspace{-2mm}
\subsection{Data-Efficiency in Fine-Tuning}
\label{sec:fine_tuning_data_eff}
We test the data-efficiency of our representations while fine-tuning, by varying the percentage of labelled data they are exposed to. We create stratified splits of the \mimic{} training set, maintaining the label distribution, with 100\%, 50\%, 25\%, 12.5\%, 10\% and 1\% of the labelled data. All smaller subset splits are members of the larger split (i.e. all images in the 1\% split are included in the 10\% split, and so on). We fine-tune our \mimic{} \best representations on each of the stratified splits from labelled \mimic{} training data. We evaluate each fine-tuned network on held-out \mimic{} validation data, and also assess zero-shot transfer on \chex{} and \VinDr{}. We observe that fine-tuning, even with as little with 10\% data, improves performance on all three datasets (\tableref{tab:data_efficiency}), indicating that the representations are data-efficient. For \chex{} evaluation, we see that even 1\% fine-tuning improves performance over zero-shot transfer by \edit{0.03} AUROC. However, the fine-tuned performance still lags that of fully-supervised \chex{}, likely due to the scale of available training labels. Interestingly, 1\% fine-tuning on \VinDr{} reduces performance, while 10+\% data improves performance. We hypothesize that this may be because the models overfits to the 1\% data split and cannot generalize to distribution shifted manifold of \VinDr{}, which has a varied label distribution (\tableref{tab:vindr_data_distribution}) than \mimic{}.

% \usepackage{booktabs}
\begin{table}[ht]
\captionsetup{skip=0pt}
\scriptsize
\caption{\small Fine-tuning data efficiency: Macro AUROC for  fine-tuning \mimic{} \best pretrained representations on three held out evaluation sets fine-tuned on stratified splits of \mimic{} training data (in \%). Results compared with zero-shot evaluations and fully supervised from scratch [FS (S)] or from ImageNet [FS (IN)] pretrained weights on their respective train sets.}
\label{tab:data_efficiency}
\begin{center}
\begin{tabular}{@{}lcccccccccc@{}}
\toprule
Eval Set & 1\% & 10\% & 12.5\% & 25\% & 50\% & 100\% & Zero-Shot & \edit{Linear Probe} & FS (S) & FS (IN) \\ \midrule
\mimic{} & 0.783 & 0.792 & 0.797 & 0.800 & 0.805 & 0.810 & \edit{NA} & \edit{0.760} & 0.710 & 0.742 \\
\chex{} & 0.679 & 0.687 & 0.690 & 0.696 & 0.701 & 0.707 & 0.649 & \edit{0.743} & 0.757 & 0.788 \\
\VinDr{} & 0.740 & 0.773 & 0.786 & 0.792 & 0.805 & 0.803 & 0.765 & \edit{0.767} & 0.668 & 0.710 \\ \bottomrule
\end{tabular}
\end{center}
\end{table}

\vspace{-1em}
\section{Conclusion}
\label{sec:conclusion}
In this work, we \edit{perform the first systematic exploration of augmentation strategies on the quality of self-supervised representations, across multiple datasets and mechanisms for chest x-rays.}
We find random resized cropping to be crucial, and adding random contrast and brightness adjustments yields powerful representations. The learned representations prove to be robust to out-of-distribution data, surpass the classification accuracy of fully supervised models for various disease labels, and even generalize to unseen conditions.

\section*{Data and Code availability}
We use publicly available, large scale chest X-ray datasets, \mimic{}, \chex{}, and \VinDr{}. Data collection details are given in Section \ref{sec:data_collection} and data preprocessing steps are outlined in Appendix \ref{app:data_processing}. We open-source all our code for all our experiments and analyses in the paper on GitHub at \url{https://github.com/StanfordMIMI/siaug}. 
%. We will publicly release the codebase upon publication.

% Acknowledgments---Will not appear in anonymized version
\midlacknowledgments{This work was supported by NIH grants R01 AR077604, R01 EB002524, R01 AR079431, K24 AR062068, and P41 EB027060; NIH contracts 75N92020C00008 and 75N92020C00021 and received computational support from Stability.AI and the Institute for Human-Centered AI at Stanford. RS received support from the Dutch Research Council, independent of this work. We would
like to acknowledge Pierre Chambon for proofreading this manuscript and members of the Stanford MIMI group, Rubin Lab and Langlotz Lab for insightful discussions.}

% \clearpage
\bibliography{paper_bibliography}

% \clearpage
\appendix
\section{Chest X-ray Dataset Distributions}
\label{app:data_dist}
Data distributions and splits used for model training and validation are outlined in \edit{Tables \ref{tab:chex_mimic_distribution} and \ref{tab:vindr_data_distribution}} below. \edit{We split both \mimic{} and \chex{} datasets into training, validation and evaluation sets. Training sets are used for self-supervised pretraining, for which the labels are not utilized. We use the same training set to train linear probes, fine-tune pretrained representations, and supervised models, while using the associated labels. The validation set is used to track metrics in this phase and for hyperparameter tuning. The held-out evaluation sets are used for all evaluations including zero-shot transfer, linear probing, and fine-tuning. These evaluation splits are unseen during training and were never used to train representations, linear probes, or to fine-tune pretrained representations. For \mimic{}, we use $n=200,000$ images for training and validation, and $n=37,962$ for evaluation. Similarly, for \chex{}, we use $n=168,660$ images for training and validation, and $n=22,367$ for evaluation.}

%%%%%%%%%%%%%%%%%%%%%%
\begin{table}[htbp]
\scriptsize
\centering
\floatconts
{tab:chex_mimic_distribution}%
{\caption{\small Baseline characteristics of \chex{} and \mimic{}.}}
{\begin{tabular}{lcccc}
\toprule
& \multicolumn{2}{c}{\mimic{}} & 
\multicolumn{2}{c}{\chex{}} \\ \midrule
Pathologies & \edit{Training \& Validation} & \edit{Evaluation} & \edit{Training \& Validation} & \edit{Evaluation} \\
\midrule
Atelectasis & 48,833 (24\%) & 9461 (25\%) & 51,892 (30\%) & 7691 (34\%) \\
Cardiomegaly & 44,206 (22\%) & 8404 (22\%) & 26,989 (16\%) & 3103 (14\%) \\
Edema & 35,440 (18\%) & 6718 (18\%) & 54,755 (33\%) & 6738 (30\%) \\
Pleural Effusion & 51,836 (26\%) & 10,306 (27\%) & 78,258 (46\%) & 8219 (37\%) \\
Pneumonia & 29,969 (15\%) & 5704 (15\%) & 18,235 (11\%) & 2421 (11\%) \\
Pneumothorax & 10,294 (5\%) & 1929 (5\%) & 18,674 (11\%) & 1727 (8\%)  \\
Rib Fracture & 4444 (2\%) & 825 (2\%) & 6914 (4\%) & 1021 (5\%) \\
No Finding & 67,239 (34\%) & 12,647 (33\%) & 14,430 (9\%) & 2544 (11\%) \\ \bottomrule
\end{tabular}}
\end{table}

%%%%%%%%%%%%%%%%%%%%%%%
% \noindent 
Splits in the VinDr datasets were formed in two ways: balanced (i.e. stratified for the included conditions) and imbalanced. \edit{We term the \VinDr{} dataset, \textit{VinDr-Imbalanced}, as it consists of a large number ``No Finding'' labels ($\sim60\%$ label prevalence). We create a separate subset of \textit{VinDr-Balanced} by undersampling this majority class to $\sim30\%$ label prevalence (all data splits in \tableref{tab:vindr_data_distribution}). We use \textit{VinDr-Balanced} for all evaluations, linear-probing and fine-tuning.}

\edit{We additionally report the performance of fine-tuning and fully supervised learning on the \textit{VinDr-Imbalanced} set in the Appendix, Table \ref{tab:vindr_balanced_imbalanced}. 
We report fully supervised performance on \textit{VinDr-Imbalanced} in the main paper for the following reasons: (a) fully supervised performance on VinDR-Imbalanced is higher than that on VinDR-Balanced (Table 7). This is expected as VinDr-Imbalanced contains more data, i.e. of the majority class, than VinDR-Balanced. (b) By showing that we outperform the best possible fully supervised performance (i.e. by using VinDR-Imbalanced) in the main paper, we also implicitly outperform fully supervised performance on VinDR-Balanced as given here in the Appendix, Table \ref{tab:vindr_balanced_imbalanced}. 
} 
% \usepackage{booktabs}
\begin{table}[ht]
\centering
\scriptsize
\caption{\small \VinDr{} splits distribution. \textbf{Bold} refers to unseen concepts.}
\label{tab:vindr_data_distribution}
\begin{tabular}{@{}lcccc@{}}
\toprule
\multicolumn{1}{c}{} & \multicolumn{2}{c}{VinDR-Balanced} & \multicolumn{2}{c}{VinDR-Imbalanced} \\ \midrule
\multicolumn{1}{c}{Pathologies} & \edit{Training} & \edit{Evaluation} & \edit{Training} & \edit{Evaluation} \\ \midrule
\textbf{Pulmonary fibrosis} & 1017 (11\%) & 217 (11\%) & 1017 (6\%) & 217 (6\%) \\
Cardiomegaly & 1817 (20\%) & 309 (16\%) & 1817 (11\%) & 309 (9\%) \\
\textbf{Pleural thickening} & 882 (10\%) & 169 (9\%) & 882 (5\%) & 169 (5\%) \\
\textbf{Lung Opacity} & 547 (6\%) & 84 (4\%) & 547 (3\%) & 84 (2\%) \\
Pleural effusion & 634 (7\%) & 111 (6\%) & 634 (4\%) & 111 (3\%) \\
Pneumonia & 471 (5\%) & 246 (12\%) & 471 (3\%) & 246 (7\%) \\
\textbf{Tuberculosis} & 482 (5\%) & 164 (8\%) & 482 (3\%) & 164 (5\%) \\
\textbf{Nodule/Mass} & 409 (4\%) & 176 (9\%) & 409 (3\%) & 176 (5\%) \\
No finding & 3000 (32\%) & 500 (25\%) & 10601 (63\%) & 2051 (58\%) \\ \bottomrule
\end{tabular}
\end{table}
%%%%%%%%%%%%%%%%%%%%%%%

% \usepackage{booktabs}
% \usepackage{multirow}
\begin{table}[ht]
\scriptsize
\centering
\caption{\small \edit{Fully supervised performance and Fine-tuning from \mimic{} \best representations performance on \textit{VinDr-Balanced} and \textit{VinDr-Imbalanced} datasplits. Numbers are AUROC. Abbreviations: FS (S), FS (IN): Fully supervised from scratch and from ImageNet, respectively, CM: Cardiomegaly, PE: Pleural Effusion, PNA: Pneumonia, PF: Pulmonay Fibrosis, PT: Pleural Thicknening, LO: Lung Opacity, TB: Tuberculosis.}}
\label{tab:vindr_balanced_imbalanced}
\begin{tabular}{@{}llcccc|ccccc|cc@{}}
\toprule
 &  & \multicolumn{4}{c|}{In-Distribution Pathologies} & \multicolumn{5}{c|}{Out-of-dsitribution Pathologies} & \multicolumn{1}{l}{} & \multicolumn{1}{l}{} \\ \midrule
\multicolumn{1}{c}{Dataset} & \multicolumn{1}{c|}{Strategy} & CM & PE & PNA & \begin{tabular}[c]{@{}c@{}}No \\ Finding\end{tabular} & PF & PT & LO & Mass & TB & \begin{tabular}[c]{@{}c@{}}Macro \\ AUROC\end{tabular} & \begin{tabular}[c]{@{}c@{}}OOD \\ AUROC\end{tabular} \\ \midrule
\multirow{3}{*}{\textit{\begin{tabular}[c]{@{}l@{}}VinDr-\\ Balanced\end{tabular}}} & \multicolumn{1}{l|}{FS (S)} & 0.796 & 0.644 & 0.610 & 0.732 & 0.601 & 0.633 & 0.616 & 0.553 & 0.648 & 0.648 & 0.610 \\
 & \multicolumn{1}{l|}{FS (IN)} & 0.871 & 0.731 & 0.659 & 0.801 & 0.643 & 0.651 & 0.565 & 0.615 & 0.680 & 0.690 & 0.631 \\
 & \multicolumn{1}{l|}{FT} & \textbf{0.937} & \textbf{0.824} & \textbf{0.790} & \textbf{0.869} & \textbf{0.719} & 0.707 & 0.660 & \textbf{0.651} & \textbf{0.802} & \textbf{0.773} & \textbf{0.708} \\ \midrule
\multirow{3}{*}{\textit{\begin{tabular}[c]{@{}l@{}}VinDr-\\ Imbalanced\end{tabular}}} & \multicolumn{1}{l|}{FS (S)} & 0.796 & 0.643 & 0.591 & 0.813 & 0.631 & 0.665 & 0.622 & 0.598 & 0.649 & 0.668 & 0.633 \\
 & \multicolumn{1}{l|}{FS (IN)} & 0.888 & \textbf{0.872} & 0.672 & 0.778 & 0.631 & 0.694 & 0.613 & 0.571 & 0.675 & 0.710 & 0.637 \\
 & \multicolumn{1}{l|}{FT} & \textbf{0.917} & 0.810 & \textbf{0.777} & \textbf{0.868} & \textbf{0.714} & \textbf{0.708} & \textbf{0.661} & \textbf{0.649} & \textbf{0.797} & \textbf{0.767} & \textbf{0.706} \\ \bottomrule
\end{tabular}
\end{table}
%%%%%%%%%%%%%%%%%%%%%%%

\section{Data Pre-Processing}
\label{app:data_processing}
Data was acquired in DICOM format for \mimic{} and \VinDr{}, while \chex{} images had been obtained as pre-processed images in JPEG format. Data was pre-processed on the basis of the DICOM headers. Images were corrected for photometric interpretation, windowed according to their respective window center and width, and scaled with an intercept and slope, if applicable. All images were resized to 224x224 pixels.

\section{Augmentations}
\label{app:appendix_aug}
The augmentations referred to in the main text were implemented using the Kornia library \cite{eriba2019kornia} for PyTorch. \edit{We explored the addition of a third augmentation from a broader set of augmentations to \best strategy and found that it did not yield higher performance. As a result, we did not explore further addition of augmentations.} Apart from random resized cropping, rotation, cutout, contrast/brightness adjustments, Gaussian noise, Gaussian blur, and Sobel filtering, we explored an additional set of less commonly used augmentations, including thin plate spline transforms \cite{eriba2019kornia}, motion blur, jigsaw puzzles \cite{noroozi2016unsupervised}, and plasma fractals \cite{nicolaou2022tormentor}. These augmentations were evaluated on an individual basis as an add-on to the augmentations of $t_\theta$. None of these augmentations surpassed the performance of $t_\theta$ \edit{while linear probing \mimic{}. We note that more complicated strategies based on our findings (e.g., tuning an augmentation policy) could potentially lead to further improvements, but would sacrifice the simplicity of the current approach. We hope to explore this in future work.} 

\subsection{Implementation Details}
\label{app:aug_imp_details}
We use \texttt{RandomResizedCrop} to construct crops with random scale of $0.2-1.0$ for pair-wise evaluation and $0.3-0.9$ for $t_\theta$, and the default parameters for aspect ratio ($3/4-4/3$). Brightness/contrast adjustment is implemented using \texttt{ColorJitter} with brightness and contrast arguments ($\lambda$) set to $0.5$ for pair-wise evaluations and $t_\theta$. A kernel size of 23 with a sigma of $0.1-2.0$ was used for Gaussian blur, whereas Gaussian noise had $\mu$ set to $0$ and $\sigma$ uniformly sampled from $0.01-0.03$. Cutout was implemented by \texttt{RandomErasing} with the default parameters (scale range $0.02-0.33$, ratio range $0.3-3.3$). All other augmentations were implemented using the default parameters as supplied by the Kornia library.

\subsection{RandAugment}
\label{app:appendix_randaugment}
The RandAugment \cite{cubuk2020randaugment} strategy was applied using all augmentations mentioned in Section \ref{app:aug_imp_details} with the number of of augmentations ($n$) set to 3, and a magnitude defined by the same hyperparameters as described above. 

\section{Additional Training Details}
\label{app:addtional_training_details}
Representations pretrained for optimal augmentation selection were trained for 50 epochs whose training duration ranged from approximately 6 to 12 hours. The corresponding linear probes were trained for 40 epochs. Checkpoints from the final epochs were used for evaluation. The \best set of augmentations was retrained for 100 epochs, and linear probes were trained for 90 epochs. Linear probes, fine-tuned models, and fully supervised models were trained free of augmentations to investigate the effectiveness of the pretrained embeddings. Fine-tuned and supervised models were trained for 90 (\mimic{} and \chex{}) and 150 (\VinDr{}) epochs.

\section{Multi-Label Metrics}
\label{app:metrics}
Evaluating multi-label classification performance is more nuanced than typical multi-class classification scenarios. Common metrics such as accuracy and AUROC might overestimate or underestimate classifier capability. As a result, we report on three multi-label metrics including AUROC, that cover three different aspects of the classifier predictions.
The \textbf{Hamming loss} is an example-based metric \cite{Tsoumakas2010_multilabel} and computes the fraction of misclassified labels across each sample and across each label. The lower the Hamming loss, the better. It is mathematically defined as $ H = \frac{1}{NK}\sum_{i=i}^n\sum_{j=1}^K [p_{ij} \neq y_{ij}]$ where $p_{ij}$ is the prediction, $y_{ij}$ is the label, $K$ is the number of classes and $N$ is the number of samples.
The \textbf{Ranking Error} \cite{Tsoumakas2010_multilabel} is a ranking type of metric that computes the number of times the irrelevant labels (i.e., low probability labels) are ranked higher than relevant labels. The lower the Ranking error, the better.

\section{Additional Results}
\label{app:add_results}

\subsection{Pairwise Augmentations}
\label{app:add_results_pairwise_aug}
\edit{We report the performance of pairwise augmentations with \mimic{} pretraining followed by linear probing using three different metrics: Macro AUROC, Hamming Loss and Ranking Error in Table \ref{tab:mimic_pairwise}. We also report the class-wise AUROC scores for each of these pairs of augmentations on \mimic{} pretraining and linear probing in Table \ref{tab:class_wise_mimic}. Similarly, performance of pairwise augmentations with \chex{} pretraining and linear probing are reported in Tables \ref{tab:chex_pairwise} and \ref{tab:class_wise_chex}.}

\subsection{Zero-shot Generalization Comparison}
\label{app:add_results_zero_shot_comp}
\edit{We show in Table \ref{tab:zero_shot_compare} that linear probes trained on \best representations that were pretrained  on \mimic{} achieve better zero-shot generalization to other datasets (\chex{} and \VinDr{}), compared to fully supervised models trained from scratch on \mimic{} as well as those initialized with ImageNet pretrained weights. This strongly suggests that augmentation-based pretraining leads to more generalizable models.}

\subsection{Generalization and Fine-tuning Results for Pairwise Augmentations}
\label{app:pairwise_augs_additional_results}

\edit{We compare various pairwise augmentations in Section \ref{sec:optimal_aug_strategy} to determine the optimal \best representations by evaluating the downstream classification performance of linear probes, trained on top of the representations. Here, we further compare the different pairs of augmentations on their generalization capabilities and fine-tuning performance. We choose a subset of these pairs (three containing Crop \& Resize and three without), namely: (a) Distort-Sobel, (b) Rotate-Distort, (c) Rotate-Sobel, (d) Crop \& Resize-Noise, (e) Sobel-Crop \& Resize, and (f) Crop \& Resize-Distort for further evaluation. We report the fine-tuning performance in Tables \ref{tab:mimic_fine_tune_pairwise} and \ref{tab:vindr_fine_tune_pairwise} and the generalization capabilties and performance of their linear probes in Table \ref{tab:mimic_lcls_gen_pairwise} and Table \ref{tab:vindr_lcls_pairwise} respectively.}

\edit{We see that augmentation pairs with Crop \& Resize outperform augmentations without Crop \& Resize consistently, following the same trend in Section \ref{sec:optimal_aug_strategy}. This strongly suggests that Crop \& Resize must be one of the augmentations in the pair. Indeed, while in majority of the cases, augmentations Crop \& Resize and Distort (which constitute the best pair of augmentations in our experiments) outperform other pairs, in some cases, the augmentation pair of Sobel and Crop \& Resize shows higher performance. However, we still believe that the combination of Crop \& Resize and Distort may be superior overall. This is because Distort can be applied in various levels, giving a degree of flexibility and control to the practitioner in selecting those levels via hyperparameter setting unlike Sobel which has limited flexibility.  }

\subsection{Generalization to Siamese Representation Learning Strategies}
\label{app:other_architectures}
\edit{Here, we evaluate whether the augmentation strategy \best  generalizes to other Siamese representation learning strategies. We consider three commonly used frameworks: SimCLR \cite{chen2020simple}, DINO \cite{caron2021emerging}, and MoCo \cite{chen2020improved}. Our findings show that \best generalizes well to a variety of pretraining strategies, including those that rely on negative pairs, such as SimCLR (\tableref{tab:other_architectures}). All strategies were trained with the default settings with a $ResNet50$ backbone for 100 epochs with a batch size of 256. The DINO framework was trained as outlined in \cite{caron2021emerging} with six local crops, and in a setting with global crops only.
}

\subsection{Linear Probing results on \chex{}}
\label{app:expanded_generalization_to_chexpert}

\edit{We expand Table \ref{tab:generalization_to_chexpert} in Section \ref{sec:transfer_to_chex}, to include Linear Probing results with \best representations on \chex{} data in Table \ref{tab:expanded_generalization_to_chex}. We see that Linear probing results in some degree of catastrophic forgetting while fine-tuning does not.}

% \usepackage{booktabs}
\begin{table}[ht]
\scriptsize
\caption{\small Pairwise Augmentations Performance with \mimic{} pretraining and linear-probing  }
\label{tab:mimic_pairwise}
\centering
\begin{tabular}{@{}llccc@{}}
\toprule
\multicolumn{1}{c}{Augmentation 1} & \multicolumn{1}{c}{Augmentation 2} & Macro AUROC $\uparrow$ & Hamming Loss $\downarrow$ & Ranking Error $\downarrow$ \\ \midrule
\multicolumn{2}{l}{Fully Supervised (Scratch)} & 0.71 & 0.186 & 0.291 \\
\multicolumn{2}{l}{Fully Supervised (ImageNet)} & 0.742 & 0.174 & 0.226 \\
Blur & Cutout & 0.676 & 0.181 & 0.224 \\
Blur & Identity & 0.671 & 0.191 & 0.237 \\
Blur & Jitter & 0.697 & 0.181 & 0.216 \\
Blur & Noise & 0.675 & 0.192 & 0.239 \\
Blur & Rotate & 0.676 & 0.192 & 0.233 \\
Blur & Crop \& Resize & 0.749 & 0.166 & 0.181 \\
Blur & Sobel & 0.686 & 0.177 & 0.216 \\
Cutout & Blur & 0.68 & 0.182 & 0.224 \\
Cutout & Identity & 0.671 & 0.187 & 0.231 \\
Cutout & Distort & 0.686 & 0.186 & 0.223 \\
Cutout & Noise & 0.675 & 0.186 & 0.23 \\
Cutout & Rotate & 0.703 & 0.182 & 0.218 \\
Cutout & Crop \& Resize & 0.747 & 0.171 & 0.189 \\
Cutout & Sobel & 0.69 & 0.176 & 0.214 \\
Identity & Identity & 0.668 & 0.187 & 0.234 \\
Distort & Blur & 0.695 & 0.181 & 0.217 \\
Distort & Cutout & 0.692 & 0.183 & 0.218 \\
Distort & Identity & 0.694 & 0.18 & 0.217 \\
Distort & Noise & 0.696 & 0.18 & 0.216 \\
Distort & Rotate & 0.709 & 0.184 & 0.215 \\
Distort & Crop \& Resize & \textbf{0.76} & \textbf{0.163} & \textbf{0.175} \\
Distort & Sobel & 0.708 & 0.174 & 0.205 \\
Noise & Blur & 0.674 & 0.189 & 0.233 \\
Noise & Cutout & 0.676 & 0.185 & 0.232 \\
Noise & Identity & 0.671 & 0.192 & 0.237 \\
Noise & Distort & 0.696 & 0.18 & 0.215 \\
Noise & Rotate & 0.673 & 0.192 & 0.24 \\
Noise & Crop \& Resize & 0.748 & 0.166 & 0.184 \\
Noise & Sobel & 0.686 & 0.175 & 0.215 \\
Rotate & Blur & 0.681 & 0.19 & 0.23 \\
Rotate & Cutout & 0.698 & 0.183 & 0.22 \\
Rotate & Identity & 0.668 & 0.196 & 0.245 \\
Rotate & Distort & 0.714 & 0.183 & 0.213 \\
Rotate & Noise & 0.674 & 0.191 & 0.237 \\
Rotate & Crop \& Resize & 0.745 & 0.168 & 0.186 \\
Rotate & Sobel & 0.705 & 0.179 & 0.21 \\
Crop \& Resize & Blur & 0.751 & 0.165 & 0.18 \\
Crop \& Resize & Cutout & 0.747 & 0.173 & 0.19 \\
Crop \& Resize & Identity & 0.745 & 0.169 & 0.187 \\
Crop \& Resize & Distort & \textbf{0.761} & \textbf{0.16} & \textbf{0.174} \\
Crop \& Resize & Noise & 0.754 & 0.165 & 0.179 \\
Crop \& Resize & Rotate & 0.748 & 0.168 & 0.184 \\
Crop \& Resize & Sobel & 0.738 & 0.167 & 0.187 \\
Sobel & Blur & 0.686 & 0.178 & 0.218 \\
Sobel & Cutout & 0.692 & 0.176 & 0.213 \\
Sobel & Identity & 0.685 & 0.176 & 0.217 \\
Sobel & Distort & 0.701 & 0.179 & 0.213 \\
Sobel & Noise & 0.684 & 0.176 & 0.217 \\
Sobel & Rotate & 0.701 & 0.182 & 0.215 \\
Sobel & Crop \& Resize & 0.755 & 0.161 & 0.174 \\ \bottomrule
\end{tabular}
\end{table}

\begin{table}[ht]
\scriptsize
\caption{\small \small Class-wise AUROC for pairwise augmentations with \mimic{} pretraining and linear-probing. Abbreviations: AT: Atelectasis, CM: Cardiomegaly, RF: Rib fracture, PE: Pleural Effusion, PNA: Pneumonia, PTX: Pneumothorax. }
\label{tab:class_wise_mimic}
\begin{tabular}{llrrrrrrrr}
\toprule
Augmentation 1 & Augmentation 2 & \multicolumn{1}{l}{AT} & \multicolumn{1}{l}{CM} & \multicolumn{1}{l}{Edema} & \multicolumn{1}{l}{RF} & \multicolumn{1}{l}{PE} & \multicolumn{1}{l}{PNA} & \multicolumn{1}{l}{PTX} & \multicolumn{1}{l}{\begin{tabular}[c]{@{}l@{}}No \\ Finding\end{tabular}} \\ \midrule
\multicolumn{2}{l}{Fully Supervised (Scratch)} & 0.705 & 0.72 & 0.797 & 0.529 & 0.837 & 0.617 & 0.689 & 0.786 \\
\multicolumn{2}{l}{Fully Supervised (ImageNet)} & 0.744 & 0.753 & 0.832 & 0.563 & 0.856 & 0.667 & 0.748 & 0.77 \\
Blur & Cutout & 0.675 & 0.694 & 0.753 & 0.544 & 0.733 & 0.571 & 0.695 & 0.742 \\
Blur & Identity & 0.662 & 0.684 & 0.763 & 0.552 & 0.731 & 0.567 & 0.678 & 0.728 \\
Blur & Distort & 0.685 & 0.706 & 0.787 & 0.569 & 0.763 & 0.581 & 0.719 & 0.764 \\
Blur & Noise & 0.666 & 0.686 & 0.766 & 0.547 & 0.733 & 0.579 & 0.687 & 0.736 \\
Blur & Rotate & 0.671 & 0.68 & 0.765 & 0.56 & 0.738 & 0.576 & 0.68 & 0.74 \\
Blur & Crop \& Resize & 0.741 & 0.766 & 0.838 & 0.599 & 0.832 & 0.642 & 0.763 & 0.812 \\
Blur & Sobel & 0.678 & 0.7 & 0.768 & 0.559 & 0.739 & 0.586 & 0.704 & 0.752 \\
Cutout & Blur & 0.675 & 0.697 & 0.761 & 0.546 & 0.74 & 0.575 & 0.699 & 0.746 \\
Cutout & Identity & 0.665 & 0.686 & 0.746 & 0.557 & 0.722 & 0.57 & 0.69 & 0.73 \\
Cutout & Distort & 0.675 & 0.701 & 0.769 & 0.56 & 0.743 & 0.583 & 0.704 & 0.754 \\
Cutout & Noise & 0.67 & 0.688 & 0.754 & 0.554 & 0.727 & 0.578 & 0.693 & 0.736 \\
Cutout & Rotate & 0.684 & 0.713 & 0.793 & 0.573 & 0.771 & 0.595 & 0.728 & 0.768 \\
Cutout & Crop \& Resize & 0.732 & 0.749 & 0.833 & 0.614 & 0.828 & 0.648 & 0.768 & 0.801 \\
Cutout & Sobel & 0.682 & 0.704 & 0.774 & 0.56 & 0.743 & 0.592 & 0.705 & 0.757 \\
Identity & Identity & 0.662 & 0.681 & 0.759 & 0.542 & 0.724 & 0.569 & 0.677 & 0.73 \\
Distort & Blur & 0.681 & 0.706 & 0.788 & 0.567 & 0.759 & 0.585 & 0.707 & 0.766 \\
Distort & Cutout & 0.677 & 0.703 & 0.779 & 0.563 & 0.748 & 0.589 & 0.714 & 0.762 \\
Distort & Identity & 0.683 & 0.703 & 0.783 & 0.563 & 0.758 & 0.587 & 0.711 & 0.766 \\
Distort & Noise & 0.683 & 0.707 & 0.788 & 0.566 & 0.762 & 0.591 & 0.705 & 0.767 \\
Distort & Rotate & 0.698 & 0.718 & 0.803 & 0.567 & 0.778 & 0.611 & 0.721 & 0.779 \\
Distort & Crop \& Resize & 0.75 & 0.769 & 0.848 & 0.619 & 0.845 & 0.649 & 0.779 & 0.819 \\
Distort & Sobel & 0.695 & 0.714 & 0.797 & 0.582 & 0.77 & 0.602 & 0.726 & 0.775 \\
Noise & Blur & 0.668 & 0.685 & 0.764 & 0.558 & 0.733 & 0.565 & 0.683 & 0.734 \\
Noise & Cutout & 0.673 & 0.687 & 0.755 & 0.565 & 0.728 & 0.577 & 0.691 & 0.735 \\ 
Noise & Identity & 0.665 & 0.684 & 0.762 & 0.54 & 0.734 & 0.574 & 0.673 & 0.735 \\
Noise & Distort & 0.681 & 0.707 & 0.788 & 0.565 & 0.761 & 0.592 & 0.707 & 0.765 \\
Noise & Rotate & 0.667 & 0.68 & 0.764 & 0.563 & 0.731 & 0.573 & 0.669 & 0.734 \\
Noise & Crop \& Resize & 0.738 & 0.765 & 0.839 & 0.615 & 0.832 & 0.634 & 0.757 & 0.807 \\
Noise & Sobel & 0.682 & 0.7 & 0.765 & 0.56 & 0.739 & 0.589 & 0.703 & 0.754 \\
Rotate & Blur & 0.672 & 0.684 & 0.766 & 0.562 & 0.745 & 0.583 & 0.689 & 0.745 \\
Rotate & Cutout & 0.682 & 0.705 & 0.785 & 0.564 & 0.769 & 0.591 & 0.729 & 0.765 \\
Rotate & Identity & 0.662 & 0.675 & 0.75 & 0.555 & 0.722 & 0.57 & 0.684 & 0.726 \\
Rotate & Distort & 0.7 & 0.727 & 0.809 & 0.565 & 0.786 & 0.617 & 0.731 & 0.779 \\
Rotate & Noise & 0.668 & 0.682 & 0.764 & 0.568 & 0.73 & 0.573 & 0.675 & 0.73 \\
Rotate & Crop \& Resize & 0.731 & 0.756 & 0.84 & 0.6 & 0.824 & 0.64 & 0.765 & 0.803 \\
Rotate & Sobel & 0.689 & 0.711 & 0.794 & 0.578 & 0.773 & 0.598 & 0.726 & 0.772 \\
Crop \& Resize & Blur & 0.741 & 0.767 & 0.841 & 0.604 & 0.833 & 0.644 & 0.767 & 0.809 \\
Crop \& Resize & Cutout & 0.734 & 0.749 & 0.835 & 0.612 & 0.828 & 0.649 & 0.768 & 0.803 \\
Crop \& Resize & Identity & 0.732 & 0.76 & 0.835 & 0.609 & 0.824 & 0.636 & 0.76 & 0.805 \\
Crop \& Resize & Distort & 0.753 & 0.771 & 0.849 & 0.608 & 0.844 & 0.655 & 0.786 & 0.821 \\
Crop \& Resize & Noise & 0.743 & 0.766 & 0.843 & 0.617 & 0.834 & 0.644 & 0.768 & 0.813 \\
Crop \& Resize & Rotate & 0.734 & 0.762 & 0.843 & 0.601 & 0.823 & 0.643 & 0.771 & 0.808 \\
Crop \& Resize & Sobel & 0.727 & 0.754 & 0.832 & 0.589 & 0.809 & 0.631 & 0.753 & 0.806 \\
Sobel & Blur & 0.678 & 0.699 & 0.764 & 0.562 & 0.74 & 0.585 & 0.708 & 0.753 \\
Sobel & Cutout & 0.684 & 0.701 & 0.778 & 0.563 & 0.749 & 0.593 & 0.71 & 0.759 \\
Sobel & Identity & 0.681 & 0.701 & 0.766 & 0.557 & 0.737 & 0.585 & 0.705 & 0.751 \\
Sobel & Distort & 0.685 & 0.707 & 0.784 & 0.584 & 0.766 & 0.595 & 0.721 & 0.771 \\
Sobel & Noise & 0.681 & 0.699 & 0.765 & 0.556 & 0.736 & 0.585 & 0.702 & 0.748 \\
Sobel & Rotate & 0.683 & 0.708 & 0.789 & 0.578 & 0.767 & 0.595 & 0.724 & 0.766 \\
Sobel & Crop \& Resize & 0.745 & 0.773 & 0.849 & 0.603 & 0.829 & 0.656 & 0.763 & 0.823 \\ \bottomrule
\end{tabular}
\end{table}

% \usepackage{booktabs}
\begin{table}[ht]
\scriptsize
\caption{\small Pairwise Augmentations Performance with \chex{} pretraining and linear-probing  }
\label{tab:chex_pairwise}
\centering
\begin{tabular}{@{}llccc@{}}
\toprule
\multicolumn{1}{c}{Augmentation 1} & \multicolumn{1}{c}{Augmentation 2} & Macro AUROC $\uparrow$ & Hamming Loss$\downarrow$ & Ranking Error $\downarrow$ \\ \midrule
\multicolumn{2}{c}{Fully Supervised (Scratch)} & 0.757 & 0.16 & 0.165 \\
\multicolumn{2}{c}{Fully Supervised  (ImageNet)} & 0.788 & 0.153 & 0.145 \\
Distort & Blur & 0.64 & 0.183 & 0.244 \\
Distort & Cutout & 0.604 & 0.187 & 0.258 \\
Distort & Noise & 0.661 & 0.181 & 0.228 \\
Distort & Rotate & 0.668 & 0.179 & 0.227 \\
Distort & Crop \& Resize & \textbf{0.736} & \textbf{0.167} & \textbf{0.189} \\
Distort & Sobel & 0.666 & 0.179 & 0.233 \\
Noise & Blur & 0.66 & 0.181 & 0.229 \\
Noise & Cutout & 0.653 & 0.182 & 0.233 \\
Noise & Noise & 0.676 & 0.179 & 0.219 \\
Noise & Rotate & 0.663 & 0.181 & 0.227 \\
Noise & Crop \& Resize & 0.691 & 0.176 & 0.214 \\
Noise & Sobel & 0.594 & 0.186 & 0.263 \\
Crop \& Resize & Blur & 0.713 & 0.172 & 0.203 \\
Crop \& Resize & Cutout & 0.73 & 0.169 & 0.193 \\
Crop \& Resize & Noise & 0.736 & 0.166 & 0.186 \\
Crop \& Resize & Rotate & 0.688 & 0.176 & 0.215 \\
Crop \& Resize & Crop \& Resize & 0.708 & 0.174 & 0.206 \\
Crop \& Resize & Sobel & 0.725 & 0.17 & 0.195 \\
Sobel & Blur & 0.631 & 0.184 & 0.248 \\
Sobel & Cutout & 0.616 & 0.187 & 0.261 \\
Sobel & Noise & 0.651 & 0.182 & 0.238 \\
Sobel & Rotate & 0.517 & 0.213 & 0.432 \\
Sobel & Crop \& Resize & 0.638 & 0.183 & 0.245 \\
Sobel & Sobel & 0.712 & 0.171 & 0.201 \\ \bottomrule
\end{tabular}
\end{table}

\begin{table}[ht]
\scriptsize
\caption{\small Class-wise AUROC for pairwise augmentations with \chex{} pretraining and linear-probing. Abbreviations: AT: Atelectasis, CM: Cardiomegaly, RF: Rib fracture, PE: Pleural Effusion, PNA: Pneumonia, PTX: Pneumothorax. }
\label{tab:class_wise_chex}
\centering
\begin{tabular}{llrrrrrrrr}
\toprule
Augmentation 1 & Augmentation 2 & \multicolumn{1}{l}{AT} & \multicolumn{1}{l}{CM} & \multicolumn{1}{l}{Edema} & \multicolumn{1}{l}{RF} & \multicolumn{1}{l}{PE} & \multicolumn{1}{l}{PNA} & \multicolumn{1}{l}{PTX} & \multicolumn{1}{l}{\begin{tabular}[c]{@{}l@{}}No \\ Finding\end{tabular}} \\ \midrule
\multicolumn{2}{l}{Fully Supervised (Scratch)} & 0.685 & 0.792 & 0.787 & 0.679 & 0.84 & 0.703 & 0.73 & 0.839 \\
\multicolumn{2}{l}{Fully Supervised (ImageNet)} & 0.703 & 0.824 & 0.816 & 0.72 & 0.862 & 0.743 & 0.783 & 0.857 \\
Distort & Blur & 0.604 & 0.622 & 0.675 & 0.623 & 0.679 & 0.6 & 0.626 & 0.69 \\
Distort & Cutout & 0.569 & 0.579 & 0.62 & 0.582 & 0.643 & 0.584 & 0.614 & 0.644 \\
Distort & Noise & 0.621 & 0.667 & 0.691 & 0.627 & 0.697 & 0.602 & 0.637 & 0.742 \\
Distort & Rotate & 0.626 & 0.649 & 0.697 & 0.646 & 0.704 & 0.628 & 0.659 & 0.733 \\
Distort & Crop \& Resize & 0.668 & 0.759 & 0.772 & 0.693 & 0.786 & 0.68 & 0.716 & 0.818 \\
Distort & Sobel & 0.612 & 0.66 & 0.714 & 0.634 & 0.708 & 0.615 & 0.654 & 0.736 \\
Noise & Blur & 0.621 & 0.672 & 0.686 & 0.627 & 0.701 & 0.596 & 0.634 & 0.74 \\
Noise & Cutout & 0.618 & 0.653 & 0.685 & 0.623 & 0.691 & 0.592 & 0.63 & 0.731 \\
Noise & Distort & 0.625 & 0.687 & 0.712 & 0.639 & 0.713 & 0.623 & 0.647 & 0.762 \\
Noise & Rotate & 0.625 & 0.669 & 0.69 & 0.636 & 0.705 & 0.597 & 0.633 & 0.749 \\
Noise & Crop \& Resize & 0.644 & 0.71 & 0.72 & 0.655 & 0.735 & 0.633 & 0.656 & 0.776 \\
Noise & Sobel & 0.574 & 0.563 & 0.605 & 0.583 & 0.631 & 0.557 & 0.594 & 0.647 \\
Crop \& Resize & Blur & 0.651 & 0.723 & 0.752 & 0.675 & 0.757 & 0.663 & 0.688 & 0.796 \\
Crop \& Resize & Cutout & 0.65 & 0.722 & 0.769 & 0.704 & 0.773 & 0.689 & 0.727 & 0.803 \\
Crop \& Resize & Distort & 0.666 & 0.76 & 0.772 & 0.682 & 0.79 & 0.682 & 0.715 & 0.82 \\
Crop \& Resize & Noise & 0.64 & 0.705 & 0.719 & 0.654 & 0.731 & 0.632 & 0.655 & 0.772 \\
Crop \& Resize & Rotate & 0.641 & 0.709 & 0.746 & 0.684 & 0.748 & 0.664 & 0.693 & 0.782 \\
Crop \& Resize & Sobel & 0.655 & 0.74 & 0.764 & 0.684 & 0.764 & 0.677 & 0.715 & 0.799 \\
Sobel & Blur & 0.587 & 0.589 & 0.66 & 0.616 & 0.668 & 0.594 & 0.646 & 0.689 \\
Sobel & Cutout & 0.579 & 0.59 & 0.628 & 0.6 & 0.649 & 0.576 & 0.652 & 0.655 \\
Sobel & Distort & 0.605 & 0.642 & 0.685 & 0.613 & 0.689 & 0.602 & 0.652 & 0.719 \\
Sobel & Noise & 0.517 & 0.518 & 0.512 & 0.5 & 0.57 & 0.506 & 0.515 & 0.501 \\
Sobel & Rotate & 0.596 & 0.607 & 0.663 & 0.626 & 0.68 & 0.592 & 0.638 & 0.703 \\
Sobel & Crop \& Resize & 0.651 & 0.729 & 0.75 & 0.662 & 0.761 & 0.653 & 0.697 & 0.796 \\ \bottomrule
\end{tabular}
\end{table}

% \usepackage{booktabs}
\begin{table}[ht]
\centering
\small
\caption{\edit{\small Comparison of zero-shot generalization capabilties of linear probes on \mimic{} pretrained \best representations (Linear Probe) to those of fully supervised models trained on \mimic{} from scratch {[}FS (S){]} or from ImageNet pretrained weights {[}FS (IN){]}}}
\label{tab:zero_shot_compare}
\begin{tabular}{@{}lcccc@{}}
\toprule
 &  & \multicolumn{3}{c}{Evaluation Data} \\ \midrule
Strategy & (Pre)Training Data & \mimic{} & \chex{} & \VinDr{} \\ \midrule
Linear Probe & \mimic{} & \textbf{0.760} & \textbf{0.649} & \textbf{0.765} \\
FS (S) & \mimic{} & 0.710 & 0.612 & 0.668 \\
FS (IN) & \mimic{} & 0.742 & 0.630 & 0.710 \\ \bottomrule
\end{tabular}
\end{table}

% \usepackage{booktabs}
\begin{table}[ht]
\small
\centering
\caption{\small \edit{Fine-tuning performance of augmentation pairs on \mimic{}. Numbers are Macro AUROCs. All models are pretrained and fine-tuned on \mimic{} data.}}
\label{tab:mimic_fine_tune_pairwise}
\begin{tabular}{@{}llccc@{}}
\toprule
\multicolumn{2}{l}{\begin{tabular}[c]{@{}l@{}}Pretrained on \mimic{}\\ Fine-Tuned on \mimic{}\end{tabular}} & \multicolumn{3}{c}{Evaluation Data} \\ \midrule
Augmentation 1 & Augmentation 2 & \mimic{} & \chex{} & \VinDr{} \\ \midrule
Distort & Sobel & 0.762 & 0.660 & 0.762 \\
Rotate & Distort & 0.765 & 0.661 & 0.758 \\
Rotate & Sobel & 0.738 & 0.646 & 0.746 \\
Crop \& Resize & Noise & 0.791 & 0.697 & 0.791 \\
Sobel & Crop \& Resize & 0.801 & 0.687 & \textbf{0.807} \\
Crop \& Resize & Distort & \textbf{0.810} & \textbf{0.707} & 0.803 \\ \bottomrule
\end{tabular}
\end{table}

\begin{table}[ht]
\scriptsize
\centering
\caption{\small \edit{Fine-tuning  performance of augmentation pairs on \VinDr{}. Models are pretrained on \mimic{} and fine-tuned on \VinDr{}. Numbers are AUROC. Abbreviations: Aug: Augmentation, CM: Cardiomegaly, PE: Pleural Effusion, PNA: Pneumonia, PF: Pulmonay Fibrosis, PT: Pleural Thicknening, LO: Lung Opacity, TB: Tuberculosis.}}
\label{tab:vindr_fine_tune_pairwise}
\begin{tabular}{ll|cccccccccc}
\toprule
\multicolumn{2}{l|}{\begin{tabular}[c]{@{}l@{}}Pretrained on \mimic{}\\ Fine-Tuned on \VinDr{}\end{tabular}} & \multicolumn{10}{c}{Evaluation Data: \VinDr{}} \\ \midrule
Aug 1 & Aug 2 & \begin{tabular}[c]{@{}c@{}}Macro \\ AUROC\end{tabular} & CM & PE & PNA & \begin{tabular}[c]{@{}c@{}}No \\ Finding\end{tabular} & PF & PT & LO & Mass & TB \\ \midrule
Distort & Sobel & 0.684 & 0.828 & 0.670 & 0.618 & 0.780 & 0.649 & 0.664 & 0.631 & 0.597 & 0.723 \\
Rotate & Distort & 0.698 & 0.853 & 0.687 & 0.705 & 0.801 & 0.638 & 0.630 & 0.651 & 0.611 & 0.711 \\
Rotate & Sobel & 0.670 & 0.824 & 0.671 & 0.630 & 0.764 & 0.631 & 0.649 & 0.606 & 0.579 & 0.678 \\
Crop \& Resize & Noise & 0.766 & 0.920 & 0.830 & 0.773 & 0.883 & 0.706 & 0.708 & 0.676 & 0.643 & 0.759 \\
Sobel & Crop \& Resize & \textbf{0.790} & \textbf{0.939} & \textbf{0.837} & \textbf{0.802} & \textbf{0.889} & \textbf{0.736} & \textbf{0.752} & \textbf{0.701} & 0.649 & \textbf{0.803} \\
Crop \& Resize & Distort & 0.773 & 0.937 & 0.824 & 0.790 & 0.869 & 0.719 & 0.707 & 0.660 & \textbf{0.651} & 0.802 \\ \bottomrule
\end{tabular}
\end{table}

% \usepackage{booktabs}
\begin{table}[ht]
\small
\centering
\caption{\small \edit{Generalization performance of linear probes on \mimic{} for different augmentation pairs. Numbers are Macro AUROCs. All models are pretrained and linearly probed on \mimic{} data.}}
\label{tab:mimic_lcls_gen_pairwise}
\begin{tabular}{@{}llccc@{}}
\toprule
\multicolumn{2}{l}{\begin{tabular}[c]{@{}l@{}}Pretrained on \mimic{}\\ Linear Probes on \mimic{}\end{tabular}} & \multicolumn{3}{c}{Evaluation Data} \\ \midrule
Augmentation 1 & Augmentation 2 & \mimic{} & \chex{} & \VinDr{} \\ \midrule
Distort & Sobel & 0.708 & 0.637 & 0.679 \\
Rotate & Distort & 0.714 & 0.635 & 0.697 \\
Rotate & Sobel & 0.705 & 0.613 & 0.655 \\
Crop \& Resize & Noise & 0.754 & \textbf{0.656} & 0.741 \\
Sobel & Crop \& Resize & 0.755 & 0.653 & 0.760 \\
Crop \& Resize & Distort & \textbf{0.761} & 0.652 & \textbf{0.765} \\ \bottomrule
\end{tabular}
\end{table}

% \usepackage{booktabs}
\begin{table}[]
\scriptsize
\centering
\caption{\small \edit{Generalization performance of linear probes on \VinDr{} for different augmentation pairs. Models are pretrained on \mimic{} and linear probes are trained on \VinDr{}. Numbers are AUROC. Abbreviations: Aug: Augmentation, CM: Cardiomegaly, PE: Pleural Effusion, PNA: Pneumonia, PF: Pulmonay Fibrosis, PT: Pleural Thicknening, LO: Lung Opacity, TB: Tuberculosis.}}
\label{tab:vindr_lcls_pairwise}
\begin{tabular}{@{}ll|cccccccccc@{}}
\toprule
\multicolumn{2}{l|}{\begin{tabular}[c]{@{}l@{}}Pretrained on \mimic{}\\ Linear Probes on \VinDr{}\end{tabular}} & \multicolumn{10}{c}{Evaluation Data: \VinDr{}} \\ \midrule
Aug 1 & Aug 2 & \begin{tabular}[c]{@{}c@{}}Macro \\ AUROC\end{tabular} & CM & PE & PNA & \begin{tabular}[c]{@{}c@{}}No \\ Finding\end{tabular} & PF & PT & LO & Mass & TB \\ \midrule
Distort & Sobel & 0.663 & 0.836 & 0.646 & 0.594 & 0.769 & 0.617 & 0.635 & 0.613 & 0.592 & 0.666 \\
Rotate & Distort & 0.675 & 0.853 & 0.649 & 0.673 & 0.775 & 0.631 & 0.613 & 0.626 & 0.578 & 0.675 \\
Rotate & Sobel & 0.638 & 0.803 & 0.667 & 0.595 & 0.690 & 0.586 & 0.600 & 0.596 & 0.569 & 0.635 \\
Crop \& Resize & Noise & 0.748 & 0.897 & 0.801 & 0.756 & 0.882 & 0.662 & 0.660 & 0.705 & 0.661 & 0.706 \\
Sobel & Crop \& Resize & 0.777 & 0.923 & 0.805 & 0.781 & \textbf{0.887} & 0.715 & \textbf{0.742} & \textbf{0.724} & 0.651 & 0.764 \\
Crop \& Resize & Distort & \textbf{0.780} & \textbf{0.924} & \textbf{0.855} & \textbf{0.802} & 0.884 & \textbf{0.719} & 0.717 & 0.679 & \textbf{0.655} & \textbf{0.788} \\ \bottomrule
\end{tabular}
\end{table}

% \usepackage{booktabs}
\begin{table}[ht]
% \captionsetup{skip=-8pt}
\centering
\small
\caption{ \small \edit{Evaluation of \best linear probing on \mimic{} with other Siamese representation learning strategies. Numbers are AUROC. DINO (global) refers to DINO with 2 global crops, and 0 local crops. DINO (local) refers to DINO with 2 global crops and 6 local crops.}}
\label{tab:other_architectures}
\begin{tabular}{@{}lcccc@{}}
\toprule
 &  & \multicolumn{3}{c}{Evaluation Data} \\ \midrule
Framework & (Pre)Training Data & \mimic{} & \chex{} & \VinDr{} \\ \midrule
SimSiam & \mimic{} & 0.760 & 0.649 & 0.765 \\
SimCLR & \mimic{} & 0.779 & 0.709 & 0.810 \\
MoCo V2 & \mimic{} & 0.785 & 0.702 & 0.781 \\
DINO (global) & \mimic{} & 0.720 & 0.606 & 0.683 \\
DINO (local) & \mimic{} & 0.726 & 0.581 & 0.678 \\
FS (S) & \mimic{} & 0.710 & 0.612 & 0.668\\
FS (IN) & \mimic{} & 0.742 & 0.630 & 0.710 \\
\bottomrule
\end{tabular}
\end{table}

\begin{table}[ht]
\scriptsize
\caption{\small \edit{Expanded table for \mimic{} \best to \chex{} transfer. Macro AUROC. Abbreviations: ZS: Zero-shot transfer, LP: Linear Probing, FT: Fine-tuning, FS: Fully Supervised, S: trained from scratch, IN: ImageNet,  Chex: \chex{}, Mimic: \mimic{}, VinDr: \VinDr{}, Eval: Evaluation.}}
\label{tab:expanded_generalization_to_chex}
\resizebox{1.1\textwidth}{!}{% <------ Don't forget this %
\begin{tabular}{ccccc|ccccc|ccccc}
\toprule
\multicolumn{5}{c|}{Eval on \chex{}} & \multicolumn{5}{c|}{Eval on \mimic{}} & \multicolumn{5}{c}{Eval on \VinDr{}} \\ \midrule
ZS & \begin{tabular}[c]{@{}c@{}}LP\\ (Chex)\end{tabular} &\begin{tabular}[c]{@{}c@{}}FT\\ (Chex)\end{tabular} & \begin{tabular}[c]{@{}c@{}}FS (S)\\ (Chex)\end{tabular} & \begin{tabular}[c]{@{}c@{}}FS (IN)\\ (Chex)\end{tabular} & ZS & \begin{tabular}[c]{@{}c@{}}LP\\ (Chex)\end{tabular} & \begin{tabular}[c]{@{}c@{}}FT\\ (Chex)\end{tabular} & \begin{tabular}[c]{@{}c@{}}FS (S)\\ (Mimic)\end{tabular} & \begin{tabular}[c]{@{}c@{}}FS (IN)\\ (Mimic)\end{tabular} & ZS & \begin{tabular}[c]{@{}c@{}}LP\\ (Chex)\end{tabular} & \begin{tabular}[c]{@{}c@{}}FT\\ (Chex)\end{tabular} & \begin{tabular}[c]{@{}c@{}}FS (S)\\ (VinDr)\end{tabular} & \begin{tabular}[c]{@{}c@{}}FS (IN)\\ (VinDr)\end{tabular} \\ \midrule
0.649 & \edit{0.743} & 0.768 & 0.757 & \textbf{0.789} & 0.760 & \edit{0.728} & \textbf{0.763} & 0.710 & 0.742 & 0.765 & \edit{0.754} & \textbf{0.810} & 0.668 & 0.719 \\ \bottomrule
\end{tabular}
}
\end{table}
\clearpage

\end{document}